\documentclass[pre,twoside,twocolumn,amsmath,amssymb,floats,superscriptaddress,10pt]{revtex4-2}

\usepackage{graphicx}
\usepackage{dcolumn}
\usepackage{bm}
\usepackage{revsymb}
\usepackage[usenames]{color}
\usepackage{subfigure}
\usepackage{color}
\usepackage{physics}
\usepackage{amsmath}
\usepackage{eufrak}
\usepackage{hyperref}
\usepackage{bbold}

\usepackage[title]{appendix}
\newcommand{\vect}[1]{\boldsymbol{#1}}

\begin{document}

\title{Heat Fluctuations in Chemically Active Systems}

\author{Jo\"el Mabillard}
\email{mabillard@pks.mpg.de}
\affiliation{Max Planck Institute for the Physics of Complex Systems, N\"othnitzer Stra\ss e 38, 01187, Dresden, Germany}

\author{Christoph A. Weber}
\email{christoph.weber@physik.uni-augsburg.de}
\affiliation{Faculty of Mathematics, Natural Sciences, and Materials Engineering: Institute of Physics, University of Augsburg, Universit\"atsstra{\ss}e\ 1, 86159 Augsburg, Germany}

\author{Frank J\"{u}licher}
\email{julicher@pks.mpg.de}
\affiliation{Max Planck Institute for the Physics of Complex Systems, N\"othnitzer Stra{\ss}e 38, 01187, Dresden, Germany}
\affiliation{Center for Systems Biology Dresden, Pfotenhauerstrasse 108, 01307 Dresden, Germany}
\affiliation{Cluster of Excellence Physics of Life, TU Dresden, 01062 Dresden, Germany}

\date{\today}

\begin{abstract}
Chemically active systems such as living cells are maintained out of thermal equilibrium due to chemical events which generate heat and lead to active fluctuations. 
A key question is to understand on which time and length scales active fluctuations dominate thermal fluctuations.
Here, we formulate a stochastic field theory with   Poisson white noise to describe the heat fluctuations which are generated by stochastic chemical events and lead to active temperature fluctuations. 
We find that on large length and time scales, active fluctuations always dominate thermal fluctuations. 
However, at intermediate length and time scales, multiple crossovers exist which highlight the different characteristics of active and thermal fluctuations. Our work provides a framework to  characterize fluctuations in active systems and reveals that local equilibrium holds at certain length and time scales.
\end{abstract}

\maketitle


\section{Introduction}

Active matter systems such as propelled particles~\cite{Marchetti2013}, 
molecular motors~\cite{Julicher1997}, active gels~\cite{Prost2015},
or active droplets~\cite{Weber2019} are driven out of thermal equilibrium by chemical processes at molecular scales. 
In such chemically active systems, a continuous flux of matter and energy  drives chemical reactions, generates mechanical forces, or induces motion of molecules and macromolecular compounds. The chemical reactions transduce chemical energy into work or movements, and also release heat into the system. This continuous supply of heat can prevent thermalization to a homogeneous temperature reflecting the non-equilibrium character of active matter.

Living cells are paradigmatic examples of active systems~\cite{Alberts2013,Howard2001,Nelson2004}. 
Active cellular processes such as cell division, cell locomotion, the expression of genes or cellular signaling processes  rely on an flux of matter and energy and the availability of chemical fuels such as adenosine triphosphate (ATP) or guanosine triphosphate (GTP) which transduce chemical energy by hydrosysis to the diphosphate forms ADP and GDP. Such processes produce entropy and maintain the cell away from thermodynamic equilibrium. They also  generate and dissipate heat. Under such non-equilibrium conditions, living cells also organize  the formation and dissolution of protein-rich condensates. Such condensates are membrane-less compartments of distinct chemical composition that play a key role for the spatial organize of cellular biochemistry~\cite{Brangwynne2009,Li2012}.
Recent work studying the  formation and dissolution of P granules in {\it{C. elegans}} embryos, suggests that the physics of phase separation governed by local thermodynamic equilibrium~\cite{Hyman2014}, provides an appropriate description of the formation of these condensates~\cite{Fritsch2021}. 
It was proposed that local equilibrium conditions hold to a good approximation at length scales of about 100nm and at microsecond time scales despite the non-equilibrium conditions inside a cell.
This raises a fundamental question for chemically active systems in general, namely whether there generally exist length and time scales for which local equilibrium applies and, if so, what determines the crossover to systems that are lacking locally well-defined thermodynamic fields.

To tackle this question, we consider active and passive heat fluctuations in a chemically active system. The active fluctuations are related to the heat input associated with stochastic chemical reaction events and are described by a stochastic field theory with  Poisson white noise. Passive fluctuations are not related to chemical events but to the stochasticity in the heat transport at local equilibrium. These fluctuations are described by a stochastic field theory with Gaussian white noise. Comparing the magnitudes and the statistical properties of both types of fluctuations, in particular, the correlation function, we identify the temporal and spatial scales at which the local equilibrium hypothesis prevails. Given a characteristic time scale, we derive an analytical expression for the maximal length scale where the passive fluctuations dominate, providing an upper bound for the volume at local equilibrium. We also investigate the properties of the stochastic field theory with  Poisson white noise developed to describe the active fluctuations. The scaling behavior of the active correlation functions and the characteristics of the noise spectrum are of particular interest due to the Poissonian character of the noise. Although living cells are used throughout this work for illustrative purposes whenever a concrete example is needed, the analysis is more general and applies to other systems where activity is generated by chemical reactions.

This work is organized as follows: in Sec.~\ref{sec:temp} we formulate the equations describing the temperature profile inside the active system in three spatial dimensions, define the active and passive fluctuations, and derive their main statistical properties.  In Sec.~\ref{sec:pft} we discuss the power spectra and higher cumulants of active temperature fluctuations considering systems of spatial dimensions $d=1,2,3$. In Sec.~\ref{sec:comp}, we compare the active and passive temperature  correlation obtained in Sec.~\ref{sec:temp} and identify the dominant contribution as a function of the time and length scales. We obtain analytical expressions for the scales where  active and passive contributions are equal, providing bounds for the time and length scales with active or passive domination.  Our concluding remarks are presented in Sec.~\ref{sec:disc}. An appendix with more details on derivations is provided at the end of the document.

\section{Temperature fluctuations in chemically active systems}
\label{sec:temp}

\subsection{Temperature dynamics and fluctuations}

In a thermodynamic system, the temperature dynamics follows from the conservation of energy. Temperature dynamics is governed by a balance equation for heat,
\begin{align}
	\rho c_{P} \partial_t T(\vect{x},t) =-\vect{\nabla}\cdot\vect{j}_Q(\vect{x},t)+\dot{Q}(\vect{x},t)\;,\label{eq:tempevol}
\end{align}
where $\rho$ is the mass density, $c_{P}$ denotes the specific heat, and $\vect{j}_Q$ is the heat current density.
Moreover,  $\dot{Q}$ corresponds to a heat source  due to the conversion, for example, of  chemical or mechanical energy into heat $Q$. In Eq.~\eqref{eq:tempevol} the dot over ${Q}$ implies a time derivative.

We are interested in heat fluctuations due to stochastic chemical events. Individual chemical events give rise to a change in reaction enthalpy $h_0$, which is released as heat.
With many  reactions of the same type taking place at positions $x_i$ and at times $t_i$, the heat source is given by
\begin{align}
	\dot{Q}(\vect{x},t) & = h_0\sum_{i=1}^{n(t,t_0)}\delta^{(3)}(\vect{x}-\vect{x}_i)\delta(t-t_i)\;,\label{eq:activesource}
\end{align}
where $n(t,t_0)$ is the number of  events  having occurred between the initial time $t_0$ and $t$. We consider, for simplicity, a Poisson distribution where chemical events occur independently with a probability of a single event at time $t$ and position $\vect{x}$ given by $\lambda(\vect{x},t)d^3\vect{x}\,dt$, where $\lambda$ is the rate per unit volume.
The average number of events in the time interval $[t_0,t]$ can thus be expressed as 
$\langle n(t,t_0)\rangle=\int_V d^3\vect{x}\int_{t_0}^t dt^\prime \lambda(\vect{x},t^\prime)$, where $V$ is the volume of the system and the angular brackets denote an ensemble average. For a heat release of a single chemical event $h_0$,  the average rate of energy released per unit volume is $h_0\lambda$.  The source term $\dot{Q}$ can be further expressed as
\begin{align}
	\dot{Q}(\vect{x},t)=h_0\lambda(\vect{x},t)+\rho c_{P} \eta_\text{A}(\vect{x},t)\;,\label{eq:Qdotbar}
\end{align}
where $\eta_\text{A}$ is a space- and time-dependent temperature noise with an average $\langle{\eta}_\text{A}\rangle=0$, which we refer to as active noise.

The heat current  $\vect{j}_Q$ in Eq.~\eqref{eq:tempevol} is driven by temperature gradients. In addition, there can be fluctuations that stem from the stochasticity of heat transport, associated with thermal conductivity. The heat current reads 
\begin{align}
	\vect{j}_Q(\vect{x},t)&= -\kappa{\vect{\nabla} T(\vect{x},t)}+\vect{\eta}_Q(\vect{x},t)\;,\label{eq:currentdensity}
\end{align}
where $\kappa$ is the thermal conductivity. At thermal equilibrium, the noise $\vect{\eta}_Q$ satisfies 
\begin{subequations}\label{eq:etaqp}
\begin{align}
	\langle \vect{\eta}_Q(\vect{x},t)\rangle&=0\;,\label{eq:etaq1p}\\
	\langle {\eta}^\alpha_Q(\vect{x},t) {\eta}^\beta_Q(\vect{x}',t')\rangle&=2k_\text{B} {T}^2 \kappa \,  \delta^{\alpha\beta}\delta^{(3)}(\vect{x}-\vect{x}')\delta(t-t')\;,\label{eq:etaq2p}
\end{align}
\end{subequations}
where the indices $\alpha$ and $\beta$ denote spatial coordinates and the variance follows from a Green-Kubo relation. Here $\vect{\eta}_Q$ describes fluctuations of heat transport.

Fluctuations in heat lead to temperature fluctuations which we define and study in the following. Combining Eqs.~\eqref{eq:Qdotbar} and~\eqref{eq:currentdensity} leads to an equation for the temperature fluctuations $\delta T \equiv T-\bar{T}$, where $\bar{T}$ denotes the average temperature.
To linear order, temperature fluctuations evolve according to
\begin{align}
	 \partial_t \delta T(\vect{x},t) =\vect{\nabla} \cdot\left(\alpha \vect{\nabla} \delta T(\vect{x},t)\right) +\eta_\text{A}(\vect{x},t) +\eta_\text{P}(\vect{x},t) \;,\label{eq:heateqdeltaT}
\end{align}
where $\alpha \equiv \kappa/\rho c_{P}$ is the thermal diffusivity and $\eta_\text{P}\equiv-\vect{\nabla}\cdot\vect{\eta}_Q/\rho c_P$ is the passive temperature noise. The average temperature $\bar{T}$ satisfies
\begin{align}
	 \partial_t \bar{T}(\vect{x},t) &=\vect{\nabla} \cdot \left( \alpha{\vect{\nabla} }{{\bar{T}}(\vect{x},t)}\right)+ \frac{h_0}{\rho c_{P} }\lambda(\vect{x},t)\;.\label{eq:eqTbar}
\end{align}

The temperature profile $\delta T=\delta T_\text{A} +\delta T_\text{P}$ is the superposition of the two contributions  $\delta T_\text{A}$ and $\delta T_\text{P}$, which stem from the active noise $\eta_\text{A}$ and passive noise $\eta_\text{P}$, respectively.  Assuming that the cross-correlation of both noises vanishes and using  constant $\alpha$ for simplicity, the equations governing the dynamics of the active or the passive fluctuations $\delta T_\text{A/P}$ can be written as
\begin{align}
	 \partial_t \delta T_\text{A/P}(\vect{x},t) =\alpha\vect{\nabla}^2  \delta T_\text{A/P}(\vect{x},t) +\eta_\text{A/P}(\vect{x},t) \;.\label{eq:heateqdeltaTAP}
\end{align}
In the remainder of this section, the active and passive  fluctuations  are investigated separately by determining the corresponding correlation functions $\left\langle{\delta T_\text{A}(\vect{x},t)}{\delta T_\text{A}(\vect{0},0)}\right\rangle$ and $\left\langle{\delta T_\text{P}(\vect{x},t)}{\delta T_\text{P}(\vect{0},0)}\right\rangle$.

\subsection{Active fluctuations} 
The active fluctuations of temperature $\delta T_\text{A}(\vect{x},t)$ are due to  active processes and the associated release of energy acting as local sources of heat. Comparing Eq.~\eqref{eq:activesource}  with Eq.~\eqref{eq:Qdotbar}, the noise ${\eta}_\text{A}$ is identified as 
\begin{align}
	{\eta}_\text{A}(\vect{x},t)&=\frac{h_0}{\rho c_{P}}\sum_{i=1}^{n(t,t_0)}\delta^{(3)}(\vect{x}-\vect{x}_i)\delta(t-t_i)-\frac{h_0 }{\rho c_{P}}\lambda(\vect{x},t)\;,\label{eq:activenoise}
\end{align}
which corresponds to a white Poisson noise~\cite{FeynmanHibbs1965,Hanggi1978A,Hanggi1979}, characterized by a zero mean and delta-correlated cumulants:
\begin{subequations}
\begin{align}
    &\langle{\eta}_\text{A}(\vect{x}_1,t_1)\rangle_\text{c}=0\;,\label{eq:PFTmean}\\
	&\langle{\eta}_\text{A}(\vect{x}_1,t_1)\dots{\eta}_\text{A}(\vect{x}_m,t_m)\rangle_\text{c}\notag\\
	&=\left(\frac{h_0}{\rho c_{P}}\right)^m\lambda(\vect{x}_1,t_1)\prod_{i=1}^{m-1}\delta^{(3)}(\vect{x}_i-\vect{x}_{i+1})\delta(t_{i}-t_{i+1})\;,\label{eq:PFTcumulants}
\end{align}
\end{subequations}
where the subscript ``{c}" denotes a cumulant. 
Note that a Poissonian-type noise with delta-correlated cumulants is ubiquitous in physical chemistry and in biophysical systems~\cite{vanKampen1983,vanKampen1992}. 

For a single chemical event ($n=1$) corresponding to a heat source occurring at $\vect{x}^\prime$ and  $t^\prime$,
the heat kernel $G(\vect{x},t|\vect{x}',t')$ is the solution of 
\begin{align}
	 &\partial_t G(\vect{x},t|\vect{x}',t') 
	 \\
	 \nonumber
	 &=\alpha \vect{\nabla}^2 G(\vect{x},t|\vect{x}',t') +\delta^{(3)}(\vect{x}-\vect{x}')\delta(t-t')\;, 
\end{align}
and is given by~\cite{Landau1987,Ozisic1993}:
\begin{align}
	G(\vect{x},t|\vect{x}',t')&=\frac{\theta(t-t')}{\left[4\pi\alpha\left|t-t'\right|\right]^{\frac{3}{2}}}\exp\left[-\frac{(\vect{x}-\vect{x}')^2}{4\alpha|t-t'|}\right]\;.\label{eq:heatkernel}
\end{align}
Recall that the heat kernel $G(\vect{x},t|\vect{x}',t')$ is formally the Green's function of the heat equation and describes the propagation of heat in the system. As the solution of an initial value problem, it breaks time-reversal invariance. 

The active temperature fluctuations $\delta T_\text{A}(\vect{x},t)$ can be expressed using the heat kernel as:
\begin{align}
	\delta T_\text{A}(\vect{x},t)=\sum_{i=1}^{n(t,t_0)} \left(\frac{h_0}{\rho c_{P}} G(\vect{x},t|\vect{x}_i,t_i) \right)-\bar{T}(\vect{x},t)\;,\label{eq:deltaTA}
\end{align}
and formally corresponds to 
the field theory of a generalized Poisson noise~\cite{Hanggi1978,Hanggi1980}.

Since the number of chemically active events 
$n(t,t_0)$ is a fluctuating variable, $\delta T_\text{A}(\vect{x},t)$ is a stochastic field. In the following, we study the statistical properties of the active temperature fluctuations $\delta T_\text{A}(\vect{x},t)$. From  Eqs.~\eqref{eq:heateqdeltaTAP} and~\eqref{eq:PFTmean}, the averaged temperature fluctuation vanishes
\begin{align}
	\langle \delta T_\text{A}(\vect{x},t)\rangle &=0\;,
\end{align}
as expected from a white Poisson noise. In Appendix~\ref{ap:gfdtA}, we calculate the moment and cumulant generating functionals for $\delta T_\text{A}$, which give the $m$-point cumulant ($m>1$):
\begin{align}
&\left\langle{\delta T_\text{A}(\vect{x}_1,t_1)}...{\delta T_\text{A}(\vect{x}_m,t_m)}\right\rangle_\text{c} \notag\\
&=\int_{t_0}^{t} dt^\prime \int_Vd^3\vect{x}^\prime\ \left(\frac{h_0}{\rho c_{P}}\right)^m \lambda(t^\prime,\vect{x}^\prime)\prod_{i=1}^{m}G(\vect{x}_i,t_i|\vect{x}^\prime,t^\prime)\;.\label{eq:mpc}
\end{align}
Since the heat kernel $G(\vect{x},t|\vect{x}',t')$ can be interpreted as a propagator between the points $\vect{x}'$ at time $t'$ and $\vect{x}$ at time $t$, the $m$-point cumulant in Eq.~\eqref{eq:mpc} is related to the probability of having all fluctuations $\delta T_\text{A}(\vect{x}_i,t_i)$ originating from a single  event at position $\vect{x}^\prime$ and time $t^\prime$. 

Considering for simplicity an infinite size system, a constant rate per unit volume and the long-time limit with $t_1,t_2\gg t_0$, we find that the second cumulant  corresponding to the two-point correlation function is given as 
\begin{align}
	&\left\langle{\delta T_\text{A}(\vect{x}_1,t_1)}{\delta T_\text{A}(\vect{x}_2,t_2)}\right\rangle_{c} \notag\\
	&= \frac{\lambda h_0^2}{8\pi\alpha\rho^2 c_{P}^2|\vect{x}_1-\vect{x}_2|}\text{Erf}\left(\frac{|\vect{x}_1-\vect{x}_2|}{\sqrt{4\alpha|t_1-t_2|}}\right)\;.\label{eq:2pta}
\end{align} 
In the limit $|\vect{x}_1-\vect{x}_2|\ll\sqrt{4\alpha|t_1-t_2|}$, the second cumulant becomes
 \begin{align}
	\left\langle{\delta T_\text{A}(\vect{x}_1,t_1)}{\delta T_\text{A}(\vect{x}_2,t_2)}\right\rangle_{c} &\simeq\frac{\lambda h_0^2}{(4\pi\alpha)^{3/2}\rho^2 c_{P}^2}\frac{1}{\sqrt{t_1-t_2}}\;,\label{eq:2pafvtl}
\end{align}
whereas for $\sqrt{4\alpha|t_1-t_2|}\ll|\vect{x}_1-\vect{x}_2|$, 
 \begin{align}
	\left\langle{\delta T_\text{A}(\vect{x}_1,t_1)}{\delta T_\text{A}(\vect{x}_2,t_2)}\right\rangle_{c} &\simeq\frac{\lambda h_0^2}{8\pi\alpha \rho^2 c_{P}^2}\frac{1}{|\vect{x}_1-\vect{x}_2|}\;.\label{eq:2ptTAlimx}
\end{align}
Note that for equal times $t_1=t_2$, the relation above is exact. 

A key finding of this work is that the two-point correlation function $\left\langle{\delta T_\text{A}(\vect{x}_1,t_1)}{\delta T_\text{A}(\vect{x}_2,t_2)}\right\rangle_{c}$ of the active fluctuations arising from Poisson-distributed chemical events follows  a power-law scaling. This can be interpreted as a critical behavior as there are correlations on all length and time scales. From the equal-time correlation function, when $t_1=t_2$ in Eq.~\eqref{eq:2ptTAlimx}, we obtain the critical exponent $\eta=0$. These anomalous fluctuations are a direct consequence of the white Poisson noise and seems to be a characteristic feature of a field theory with stochastic Poisson noise.

\subsection{Passive fluctuations}

Even in the absence of active processes, there are fluctuations around the equilibrium temperature~\cite{Landau1985}. In a system of finite volume, for example, the relaxation towards equilibrium leads to an uncertainty on the actual value of $T(\vect{x},t)$ with respect to the equilibrium temperature of the system. Similarly, in the case of local equilibrium, the temperature is fixed in each volume element with a certain uncertainty.
In addition, the stochasticity of heat transport  lead to fluctuations in temperature with an amplitude that depends on the thermal conductivity. These fluctuations enter the heat equation through the Gaussian white noise $\eta_\text{P}$.
Using $\eta_\text{P}=-\vect{\nabla}\cdot\vect{\eta}_Q/(\rho c_P)$ and Eqs.~\eqref{eq:etaqp}, the passive noise satisfies
\begin{subequations}
\begin{align}
	\left\langle{\eta}_\text{P}(\vect{x},t)\right\rangle&=0\;, \\
	\left\langle{\eta}_\text{P}(\vect{x},t){\eta}_\text{P}(\vect{x}',t')\right\rangle&=-\frac{2k_B\bar{T}^2\alpha}{\rho c_{P}}\vect{\nabla}_{\vect{x}}^2\delta^{(3)}(\vect{x}-\vect{x}')\delta(t-t')\;.
\end{align}
\end{subequations}

Now we study the statistics of  the passive fluctuations of the temperature $\delta T_\text{P}$ governed by Eq.~\eqref{eq:heateqdeltaTAP}. Due to the Gaussian character of the noise, the only non-vanishing cumulant is the two-point correlation which can be derived using Fourier transformations. Due to the independence of active and passive noise,
 Eq.~\eqref{eq:heateqdeltaTAP} in Fourier space becomes
\begin{align}
	-i\omega\delta\tilde{T}_\text{P}(\vect{q},\omega)&=-\alpha q^2\delta\tilde{T}_\text{P}(\vect{q},\omega)+\tilde{\eta}_\text{P}(\vect{q},\omega)\;,
\end{align}
when using the definition of the Fourier transform
\begin{align}
	 \tilde{f}(\vect{q},\omega)&\equiv\int_{t_0}^tdt^\prime\int_Vd^3\vect{x}\ f(\vect{x},t^\prime)e^{-i(\vect{q}\cdot\vect{x}-\omega t^\prime)}\;.\label{eq:FourierT}
\end{align}
 The passive noise in Fourier space $\tilde{\eta}_\text{P}$ satisfies
\begin{subequations}
\begin{align}
	&\left\langle\tilde{\eta}_\text{P}(\vect{q},\omega)\right\rangle=0\;, \\
	&\left\langle\tilde{\eta}_\text{P}(\vect{q},\omega)\tilde{\eta}_\text{P}({\vect{q}'},\omega')\right\rangle\\
	\nonumber
	& =\frac{2k_B\bar{T}^2\alpha}{\rho c_{P}}q^2(2\pi)^4\delta^{(3)}(\vect{q}+\vect{q}')\delta(\omega+\omega')\;.
\end{align}
\end{subequations}
The two-point correlation function for passive fluctuations in Fourier space reads~\cite{Landau1980,Forster1975}:
\begin{align}
	&\left\langle{\delta \tilde{T}_\text{P}(\vect{q}_1,\omega_1)}{\delta  \tilde{T}_\text{P}(\vect{q}_2,\omega_2)}\right\rangle_\text{c}\notag\\
	&=\frac{2k_B\bar{T}^2\alpha}{\rho c_{P}}q_1^2\frac{(2\pi)^4\delta^{(3)}(\vect{q}_1+\vect{q}_2)\delta(\omega_1+\omega_2)}{\left(\alpha q_1^2-i\omega_1\right)\left(\alpha q_2^2-i\omega_2\right)}\;,
\end{align}
and taking the inverse Fourier transforms gives
\begin{align}
	&\left\langle{\delta T_\text{P}(\vect{x}_1,t_1)}{\delta T_\text{P}(\vect{x}_2,t_2)}\right\rangle_\text{c} \notag\\	
	&=\frac{k_B\bar{T}^2}{ \rho c_{P}\left(4\alpha\pi|t_1-t_2|\right)^{\frac{3}{2}}}\exp\left[-\frac{|\vect{x}_1-\vect{x}_2|^2}{4{\alpha|t_1-t_2|}}\right]\;.\label{eq:2ptp}
\end{align}
Contrary to the active fluctuations, the correlation function for the fluctuations around equilibrium does not possess any power-law scaling. In particular, if the time difference $|t-t'|$ is fixed, the two-point function decays exponentially with the distance $|\vect{x}-\vect{x}'|$.
In the limit $|\vect{x}_1-\vect{x}_2|\ll\sqrt{4\alpha|t_1-t_2|}$, the two-point correlations follow 
 \begin{align}
	\left\langle{\delta T_\text{P}(\vect{x}_1,t_1)}{\delta T_\text{P}(\vect{x}_2,t_2)}\right\rangle_{c} &\simeq\frac{k_B\bar{T}^2}{ \rho c_{P}}\frac{1}{\left(4\alpha\pi|t_1-t_2|\right)^{\frac{3}{2}}}\;,\label{eq:2ppfvtl}
\end{align}
whereas for the equal-time correlations, we obtain 
 \begin{align}
	\left\langle{\delta T_\text{P}(\vect{x}_1,t)}{\delta T_\text{P}(\vect{x}_2,t)}\right\rangle_{c} &=\frac{k_B\bar{T}^2}{ \rho c_{P}}\delta^{(3)}(\vect{x}_1-\vect{x}_2)\;.\label{eq:2ptTBlimx}
\end{align}

\section{Power spectra and higher cumulants of active temperature fluctuations}
\label{sec:pft}

The formalism developed in Sec.~\ref{sec:temp} to describe the active fluctuations is a free Poissonian stochastic field theory with  Poisson white noise $\eta_\text{A}$ given by Eq.~\eqref{eq:activenoise} which has vanishing mean and cumulants given in Eqs.~\eqref{eq:PFTmean} and \eqref{eq:PFTcumulants}. In this section, we discuss key features of the  Poisson field theory and highlight differences to a Gaussian stochastic field~\cite{ZinnJustin2002}. For the sake of generality, we consider in this section a $d$-dimensional space.

The first and the second cumulants are identical to those of  Gaussian white noise, but higher-order cumulants, given in Eq.~\eqref{eq:PFTcumulants}, are non-vanishing and delta-correlated. The latter lead to non-trivial, higher-order cumulants of the field $\delta T_\text{A}$, which are known exactly in Fourier space as 
\begin{align}
&\left\langle{\delta \tilde{T}_\text{A}(\vect{q}_1,\omega_1)}...{\delta \tilde{T}_\text{A}(\vect{q}_m,\omega_m)}\right\rangle_\text{c}\notag\\
&=\lambda \left(\frac{h_0}{\rho c_{P}}\right)^m  (2\pi)^{(d+1)}\frac{\delta^{(d)}(\vect{q}_1+...+\vect{q}_m)\delta(\omega_1+...+\omega_m)}{\left(\alpha q_1^2-i\omega_1\right)\dots\left(\alpha q_m^2-i\omega_m\right)}\;,\label{eq:npointcumulantfourierspace}
\end{align}
with the Fourier transform defined in Eq.~\eqref{eq:FourierT}. 
The non-vanishing higher-order cumulants in   Eq.~\eqref{eq:npointcumulantfourierspace} characterizes the non-Gaussian character of the Poisson field theory.

{
\begin{figure}[ht]\centering
    {\includegraphics[width=0.48\textwidth]{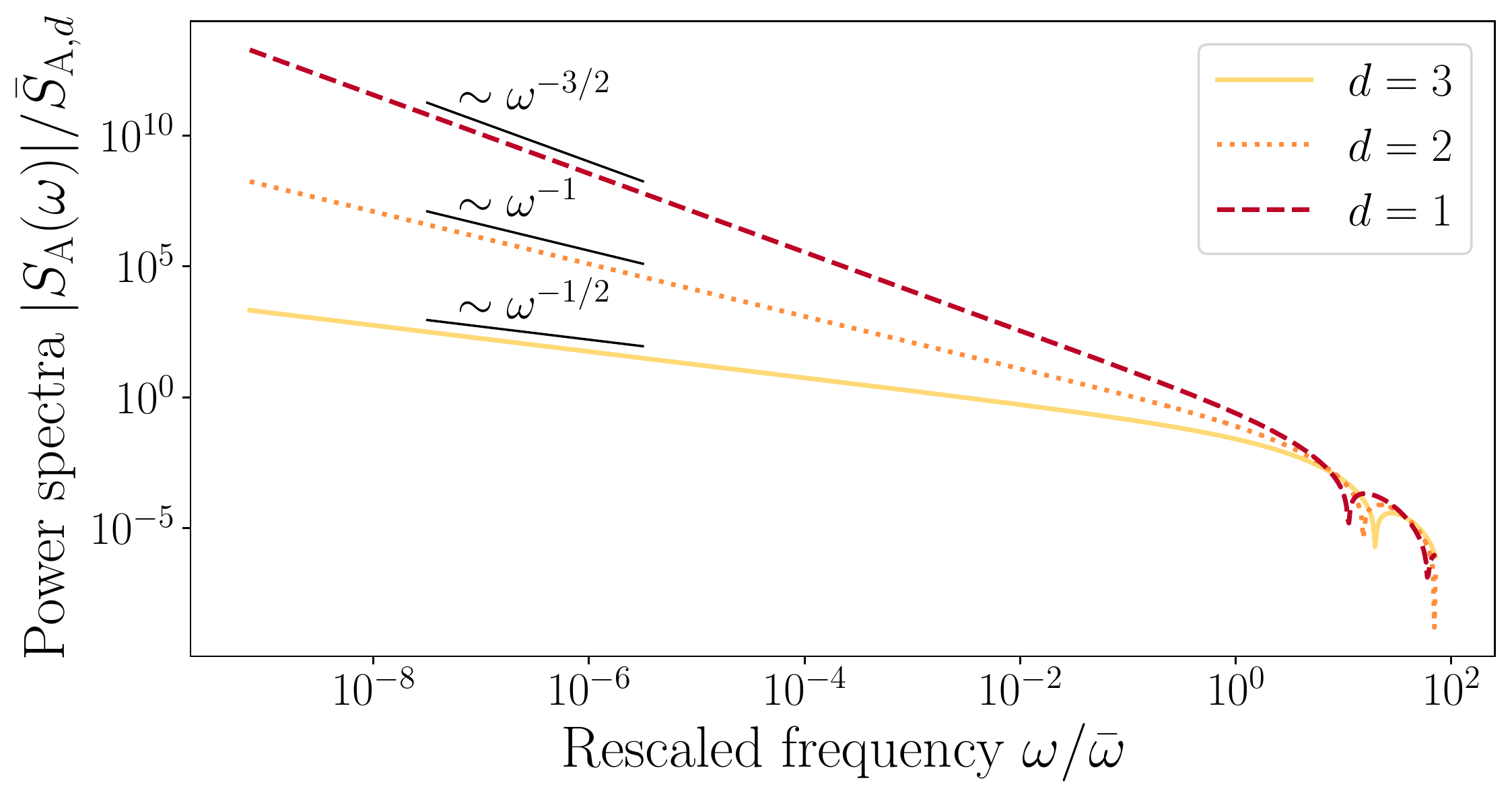}}
	\caption[]{
Power spectra $|S_{\text{A}}|$   defined in Eq.~\eqref{eq:defspect_dens} normalized by $\bar{S}_{\text{A},d}\equiv \lambda \left(\frac{h_0}{\alpha \rho c_{P}}\right)^2|\vect{x}|^{4-d}$ as function of the rescaled frequency $\omega/\bar{\omega}$ with $\bar{\omega}=\alpha/|\vect{x}|^2$ for spatial dimensions $d=1,2,3$. 
The scaling behavior $\omega^{\frac{d}{2}-2}$ is indicated. 
	}
        \label{fig:spectraldensities}
\end{figure}
}
We have shown in the previous section that the equal-time correlation function of active temperature fluctuations exhibits as a power-law scaling while for passive fluctuations it is a delta function. To explore this power-law behavior further, we investigate the power spectra of the active temperature fluctuations for systems in $1$, $2$, and $3$ spatial dimensions. We define the spectral density $S_{\text{A}}(|\vect{x}|,\omega)$ as follows~\cite{Risken1996}:
\begin{align}
	&\left\langle{\delta \hat{T}_\text{A}(\vect{x}_1,\omega_1)} \, {\delta \hat{T}_\text{A}(\vect{x}_2,\omega_2)}\right\rangle_\text{c}\notag\\
	&=2\pi \, S_{\text{A}}(|\vect{x}_1-\vect{x}_2|,\omega_1)\, \delta(\omega_1+\omega_2)\;,\label{eq:defspect_dens}
\end{align}
where $\delta\hat{T}$ is the Fourier transform of $\delta T$ in frequency space. The explicit expressions of the spectral densities are given in Eqs.~\eqref{eq:spectal_densities3}-\eqref{eq:spectal_densities1}.
In the limit of small frequencies, we find that  the  spectral  density scales as $\omega^{\frac{d}{2}-2}$,  see Fig.~\ref{fig:spectraldensities} and Eq.~\eqref{eq:spectral_density_scaling}. Note that for $d=2$, we have $S_A\sim \omega^{-1}$ which  is a $1/f$-noise~\cite{Voss1976,Schlesinger1987,Wentian1989}. Such type of noise is typical in biophysical systems~\cite{Szendro2001} and a common feature associated with Poisson shot noise~\cite{Butz1972}.

\section{Active versus passive fluctuations}
\label{sec:comp}

We are interested in length and time scales for which either active or passive fluctuations dominate. To this end, we consider  temperature correlation functions which have contributions from passive and active fluctuations. At length scales for which passive fluctuations dominate, local thermodynamics equilibrium is a valid approximation. On the contrary, for length scales where active fluctuations dominate,  local equilibrium condition is not satisfied.

The two-point temperature correlation function \begin{align}
    \mathcal{C}({x},t)= \mathcal{C}_\text{A}({x},t)+ \mathcal{C}_\text{P}({x},t)
\end{align}
is the sum of the corresponding correlation functions related to active and passive fluctuations,
\begin{align}
    \mathcal{C}_\text{A/P}({x},t)\equiv 	\left\langle{\delta T_\text{A/P}(\vect{x},t)}{\delta T_\text{A/P}(\vect{0},0)}\right\rangle_{c}\;.\label{eq:CAP}
\end{align}
Note that the correlation function $\mathcal{C}$ is defined here in terms of the cumulants which are identical to the second moments as the mean fluctuations vanish.
According to Eqs.~\eqref{eq:2pta} and~\eqref{eq:2ptp},
the correlation functions $\mathcal{C}_\text{A/P}(x,t)$ depend on $x\equiv|\vect{x}|$.
From the same equations, we find for the two-point temperature correlation function:
\begin{align}
  \mathcal{C}(x,t)
	=&\frac{\lambda h_0^2}{8\pi\alpha\rho^2 c_{P}^2x}\text{Erf}\left(\frac{x}{\sqrt{4\alpha t}}\right)\notag\\
	&+\frac{k_B\bar{T}^2}{ \rho c_{P}\left(4\alpha\pi t\right)^{\frac{3}{2}}}\exp\left[-\frac{x^2}{4{\alpha t}}\right]\;.\label{eq:full2pt}
\end{align}

The  correlation functions $ \mathcal{C}_\text{A}(x,t)$ and $ \mathcal{C}_\text{P}(x,t)$ of passive and active fluctuations  are shown on Fig.~\ref{fig:2ptfunctionactivepassive}(a-f). Fig.~\ref{fig:2ptfunctionactivepassive}(a-c) depict the active and passive temperature correlations as a function of the length scale $x$ for fixed time scales $t$. On length scales larger than the diffusion length of passive fluctuations, i.e. $x\gg\sqrt{4\alpha t}$, the passive correlations $ \mathcal{C}_\text{P}(x,t)$ are exponentially suppressed (Eq.~\eqref{eq:2ptp}), whereas the active $\mathcal{C}_\text{A}(x,t)$ correlations are independent of time $t$ and decrease as a power-law (Eq.~\eqref{eq:2ptTAlimx}). For $x\ll\sqrt{4\alpha t}$, the active and passive two-point functions, given by Eqs.~\eqref{eq:2pafvtl} and ~\eqref{eq:2ppfvtl}, respectively, reach finite values. For the time scale 
\begin{align}
	{\tau} =\frac{k_B \bar{T}^2\rho c_{P}}{\lambda h_0^2}\;,\label{eq:tauAtoP}
\end{align}
these two values are equal, such that $C_\text{A}(0,\tau)=C_\text{P}(0,\tau)$, Fig.~\ref{fig:2ptfunctionactivepassive}(b). 

{
\begin{figure}[bt]\centering
    {\includegraphics[width=0.48\textwidth]{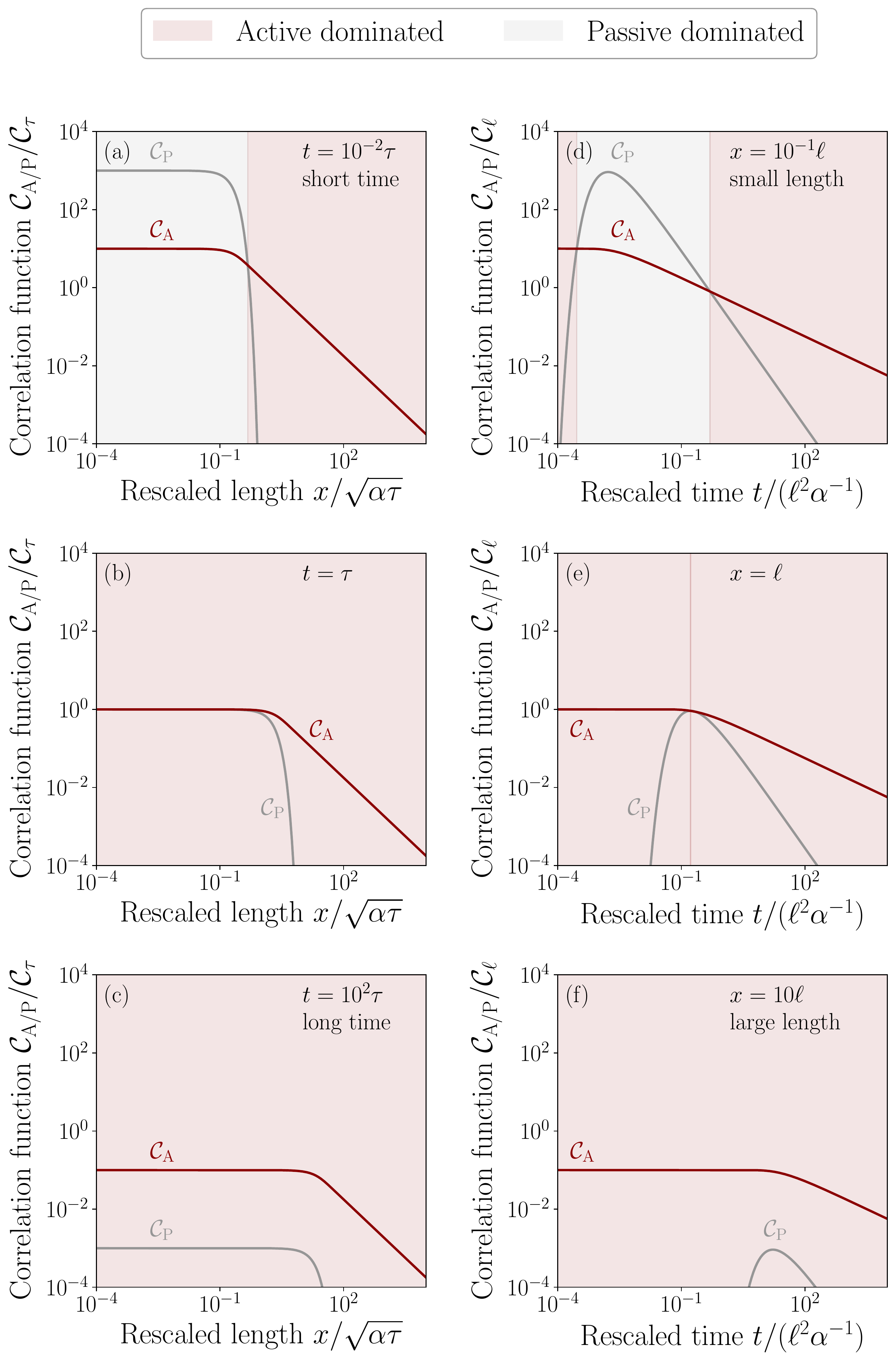}}
	\caption[]{Normalized two-point correlation functions ${\mathcal{C}_\text{P}}(x,t)$ (grey) and ${\mathcal{C}_\text{A}}(x,t)$ (red) for passive and active fluctuations. Panels (a-c): Temperature correlations $\mathcal{C}_\text{A/P}$ normalized by $\mathcal{C}_{\tau}$ where $\mathcal{C}_{\tau}= \mathcal{C}_\text{A}(0,\tau)=\mathcal{C}_\text{P}(0,\tau)$ with $\tau$ given in Eq.~\eqref{eq:tauAtoP}, as function of normalized distance $x/\sqrt{\alpha \tau}$ for $t=10^{-2}\tau$ (a), $t=\tau$ (b)  and $t=10^2\tau$ (c). Panels (d-f):  Temperature correlation  $\mathcal{C}_\text{A/P}$ normalized by $\mathcal{C}_{\ell}$ where $\mathcal{C}_{\ell}=\mathcal{C}_\text{A}(\ell,\ell^2/6\alpha)=\mathcal{C}_\text{P}(\ell,\ell^2/6\alpha)$ with $\ell$ given in Eq.~\eqref{eq:lmaxatop}, as function of normalized time $t/(\ell^2\alpha^{-1})$  for   $x=10^{-1} \ell $ (d),  $x=\ell $ (e) and $x=10 \ell $ (f). The background shade indicates the dominant contribution to the correlation function:  active (red), passive (grey). For the vertical line on panel (e), the contributions are equal. }
   \label{fig:2ptfunctionactivepassive}
\end{figure}
}

 For time scales smaller than $\tau$, Fig.~\ref{fig:2ptfunctionactivepassive}(a), the passive fluctuations dominate the temperature correlation function in Eq.~\eqref{eq:full2pt} on length scales smaller than the crossover length
\begin{align}
	L_\text{co}(t)&\simeq\left[-2\alpha t W_{-1}\left(-\frac{\lambda^2 k^2_B\pi(h_0/k_B\bar{T})^4}{2\rho^2 c_{P}^2}t^2\right)\right]^{\frac{1}{2}}\label{eq:bdactopass} \, ,
\end{align}
where $W_{-1}$ is the $-1$ branch of the Lambert $W$ function~\cite{Corless1996}. To obtain $L_\text{co}$, we have used that the active and passive correlations are equal on a length scale  that is larger than $\sqrt{4\alpha t}$ and we used Eq.~\eqref{eq:2ptTAlimx} as an  approximation for the correlation function $C_{\text{A}}$ defined in Eq.~\eqref{eq:CAP}. Note that the length $L_{\text{co}}(t)$  exists in a range of time scales $t$ that corresponds to the domain of the Lambert function $W_{-1}$. At $x=L_\text{co}$,  active and passive correlations are equal and
for length scales larger than  $L_\text{co}$, the active fluctuations dominate the two-point function of Eq.~\eqref{eq:full2pt}. For time scales larger than $\tau$, Fig.~\ref{fig:2ptfunctionactivepassive}(c), the active fluctuations dominate the passive contribution on all length scales. 

 Fig.~\ref{fig:2ptfunctionactivepassive}(d-f) show the correlation functions  $\mathcal{C}_\text{P}(x,t)$ and $\mathcal{C}_\text{A}(x,t)$ as function of the time scale $t$ for fixed values of $x$. On time scales $t\gg x^2/4\alpha$, the active and passive correlations scale in time as  $\mathcal{C}_\text{A}\sim t^{-\frac{1}{2}}$ and $\mathcal{C}_\text{P}\sim t^{-\frac{3}{2}}$ as shown in Eqs.~\eqref{eq:2pafvtl} and \eqref{eq:2ppfvtl}, respectively. The passive fluctuations have a maximum at  $t=x^2/\left(6\alpha\right)$, whereas from Eq.~\eqref{eq:2ptTAlimx} the active fluctuations reach their maximal values at $t=0$. The length scale for which the maximum of $\mathcal{C}_\text{P}(x,t)$ equals $\mathcal{C}_\text{A}(x,t)$, Fig.~\ref{fig:2ptfunctionactivepassive}(e) {(vertical line)},  is given by
\begin{align}		 \ell&=\left(\frac{6^{\frac{3}{2}}\alpha \rho c_{P}}{\pi^{\frac{1}{2}}e^{\frac{3}{2}}\text{Erf}\left(\sqrt{\frac{3}{2}}\right)k_B\, \lambda (h_0/k_B\bar{T})^2}\right)^{1/2}\;,\label{eq:lmaxatop}
\end{align}
such that $C_\text{A}(\ell,\ell^2/6\alpha)=C_\text{P}(\ell,\ell^2/6\alpha)$.
For length scales smaller than $\ell$, Fig.~\ref{fig:2ptfunctionactivepassive}(d), the passive contribution dominates for time scales in the range $t_\text{co}<t<\tau$ with
\begin{align}
	t_\text{co}(x)&\simeq-\frac{x^2}{6\alpha W_{-1}\left(-\frac{2}{3}\left[\frac{\lambda(h_0/k_B\bar{T})^2k_B\sqrt{\pi}}{8\alpha \rho c_{P}}x^2\right]^{\frac{2}{3}}\right)}\;,\label{eq:timeacttopas}
\end{align} 
where $W_{-1}$ is the $-1$ branch of the Lambert $W$ function. To obtain $t_\text{co}$, we have used that the active and passive correlations are equal on a time scale that is smaller than $x^2/4\alpha$ and we used Eq.~\eqref{eq:2ptTAlimx} as an approximation for the correlation function $C_{\text{A}}$ defined in  Eq.~\eqref{eq:CAP}.  Note that  the time $t_{\text{co}}(x)$ exists in a range of length scales $x$ that corresponds to the domain of the Lambert function $W_{-1}$. Furthermore $t_\text{co}(x)$ is the inverse function of $L_\text{co}(t)$ given in Eq.~\eqref{eq:bdactopass}. Finally, for length scales larger than $\ell$, Fig.~\ref{fig:2ptfunctionactivepassive}(f), the active fluctuations are larger than the passive contributions for all times.

\begin{figure}[t]\centering
    {\includegraphics[width=0.48\textwidth]{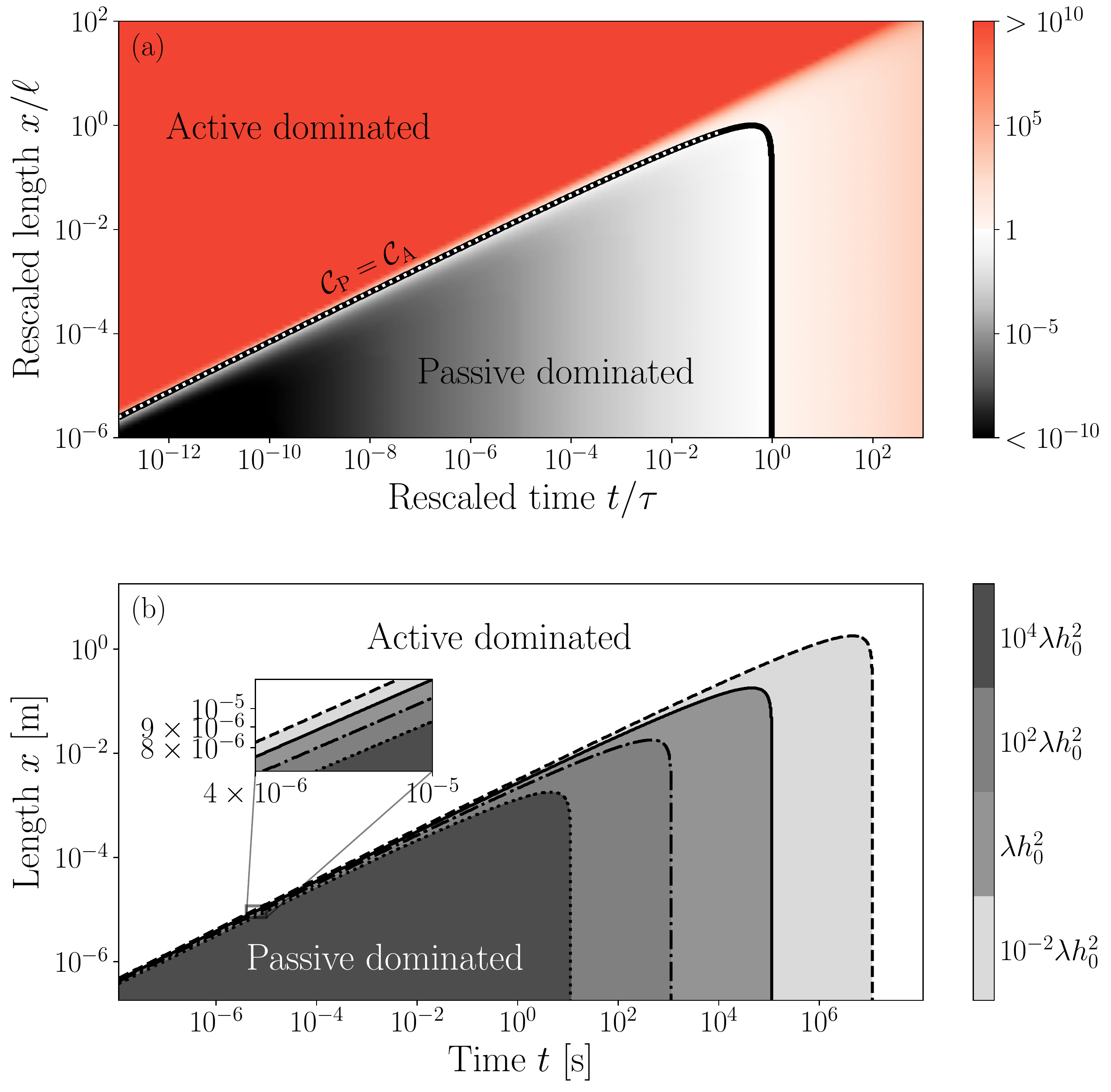}}
	\caption[]{
	 Panel (a): Active (red) or passive (grey) dominant contribution to the two-point correlation function $\mathcal{C}$, Eq.~\eqref{eq:full2pt}, in the $(t,x)$ plane as function of the renormalized time and length scales $t/\tau$ and $x/\ell$ with $\tau$ given in Eq.~\eqref{eq:bdactopass} and $\ell$ given in Eq.~\eqref{eq:lmaxatop}. The solid line depicts the boundary where $\mathcal{C}_{\text{A}}=\mathcal{C}_{\text{P}}$ between the active and passive dominated regions. The white dots on the solid boundary line indicate the crossover length $L_\text{co}(t)$  given in Eq.~\eqref{eq:bdactopass}.  The intensity of the colors  describes the ratio $\mathcal{C}_\text{A}/\mathcal{C}_\text{P}$. Panel (b): Regions  (shaded, overlaying) where passive fluctuations dominate in the $(t,x)$ plane as a function of time scale $t$ and length scale $x$.  The solid line is the limit between active and passive regions for the parameters values $h_{0}=10 k_B\bar{T}$ with $\bar{T}=289 \textrm{K}$ and $\lambda=2.5\cdot 10^4/\text{s}(\mu \text{m})^3$. This can be compared to the corresponding regions when the strength of the noise characterized by $\lambda h_0^2$ is varied by a factor $10^{-2}$ (dashed), $10^2$ (dot-dashed) or $10^4$ (dotted). Note that darker regions partially cover lighter ones.  Other parameters are $\alpha=1.4\cdot 10^{-7}\text{m}^2/\text{s}$, $\rho=10^3 \text{kg}/\text{m}^3$ and $c_{P}=4\cdot 10^3 \text{J}/(\text{kg K})$.}
        \label{fig:activevspassive}
\end{figure}

The analysis of the active and passive  two-point correlations allows us identifying the regions in the temporal and spatial scales in the $(t,x)$ plane which are dominated by either active or passive fluctuations, respectively;
see Fig.~\ref{fig:activevspassive}(a). 
The solid line corresponds to values of $x$ and $t$ for which both contributions are equal. The region dominated by passive fluctuations is bounded by $\tau$ of Eq.~\eqref{eq:tauAtoP} in the direction of increasing time scales $t$ and by $L_\text{co}$ of Eq.~\eqref{eq:bdactopass} in the direction of increasing length scale (see dotted line). 
The largest possible length scale with dominant passive fluctuations is given by  $\ell$ of Eq.~\eqref{eq:lmaxatop} and is the maximum of the solid line depicting the boundary between active and passive dominated regions. 

On large length or time scales, larger than $\ell$ and $\tau$ respectively, the system is dominated by active fluctuations. On the contrary, inside the region bounded by $L_\text{co}$ in the $x$-direction and $\tau$ in the $t$-direction, passive fluctuations dominate and equilibrium is a good approximation. To quantify the relative importance of the passive versus the active contribution,
we consider the ratio  $\mathcal{C}_\text{A}/\mathcal{C}_\text{P}$. In the active region, for time scales smaller than $\tau$ and for length scales larger than $L_\text{co}$, the passive correlation vanishes exponentially with increasing $x$ and becomes much smaller 
than the active contribution already for length scales of one order of magnitude larger than $L_\text{co}$, as seen on Fig.~\ref{fig:activevspassive}(a). In the passive dominated region, $\mathcal{C}_\text{P}$ is at least $2$, $4$ and $6$ orders of magnitude larger than $\mathcal{C}_\text{A}$ for length scales smaller than $L_\text{co}$ and values of $t$ equal to $\tau/10^2$,  $\tau/10^4$ and $\tau/10^6$ respectively. This observation allows identifying the region where the local-equilibrium hypothesis holds, i.e. where the passive correlation function strongly dominates over the active contribution.

For a given chemically active system, our analysis allows us identifying the length and time scales where local thermodynamic equilibrium is an appropriate description.
In the following, we consider the example of biological cells. For the biophysics of P granule assembly and disassembly, it has been suggested  that local thermodynamics holds on length scales of $100\text{nm}$ and time scales of tens of nanoseconds (Ref.~\cite{Fritsch2021}).  Taking the hydrolysis of ATP as a prototypical chemical event in the cell, we estimate a reaction enthalpy  $h_{0}=10 k_B\bar{T}$. Furthermore, we use $\bar{T}=289\text{K}$ and estimate the rate of events per unit volume as $\lambda\simeq 2.5\cdot 10^4/\text{s}(\mu \text{m})^3$, such that the  heat production per by unit volume and time is $\langle\dot{Q}\rangle=10^3 \text{W/m}^3$, consistent with estimates for the heat production of living tissues~\cite{bionumber2022,Thommen2019}. Considering a thermal diffusivity for the cell  similar to water $\alpha=1.4\cdot 10^{-7}\text{m}^2/\text{s}$~\cite{Blumm2003} and using the mass density  $\rho=10^3 \text{kg}/\text{m}^3$ and the specific heat $c_{P}=4\cdot 10^3 \text{J}/(\text{kg K})$ of water, the spatio-temporal scales dominated by passive fluctuations are inside a region, see  Fig.~\ref{fig:activevspassive}(b) (grey region bounded by solid line). In particular, the passive region exists for times shorter than $\tau=1.2\cdot 10^5 \text{s}$ and lengths smaller than $\ell=0.18 \text{m}$ respectively. In Ref.~\cite{Fritsch2021}, it has been observed that  $10^{-8}\text{s}$ after a local metabolic event releasing heat $h_0$, this heat has spread in a volume of linear dimension of about $10^{-7}\text{m}$. The corresponding temperature change due to the event is about $10^{-5}\text{K}$ and, thus small. Our analysis shows that for  time scales of $10^{-8}\text{s}$, passive fluctuations dominate up to length scales given by the crossover length $L_{\text{co}}(10^{-8}\text{s})=4.2\times 10^{-7}\text{m}$. This demonstrates that our analysis shown in Fig.~\ref{fig:activevspassive}(a-b) is consistent with estimates given in Ref.~\cite{Fritsch2021}.

In Fig.~\ref{fig:activevspassive}(a), the ratio between the active and passive correlation functions, $\mathcal{C}_\text{A}/\mathcal{C}_\text{P}$, has been defined as a measure of the chemical activity. The ratio depends linearly on the rate of events per unit volume $\lambda$ and quadratically on the amount of heat released by individuals events $h_0$. Fig.~\ref{fig:activevspassive}(b) shows the region of the $(t,x)$-plane dominated by passive fluctuations for different values of $h_0^2\lambda$.
Taking the values $h_{0}$ and $\lambda$ estimated above for the living cell as reference (solid line), we  observe that increasing (decreasing) $\lambda h_0^2$ decreases  (increases) the area of the passive region.
The decrease is mostly due to a decrease of the rightmost boundary along the $t$-direction which can also be seen in Eq.~\eqref{eq:tauAtoP} showing that $\tau$ is inversely proportional to $\lambda h_0^2$. 
However, the boundary in the $x$-direction does not change significantly when increasing $\lambda h_0^2$, as depicted in the inset of Fig.~\ref{fig:activevspassive}(b).
The reason for this behavior is that for time scales smaller than $\tau$, passive fluctuations always dominate  on the diagonal $x=\sqrt{4\alpha t}$,  corresponding to the diffusion length of the passive temperature fluctuations, independently of $\lambda$ and $h_0$. 
The trend of a decreasing area where passive fluctuations dominate 
confirms our expectation that the higher the rate and energy released by chemical events, the greater the activity of the system.

\section{Discussion}
\label{sec:disc}


Our analysis has been motivated by quantitative studies of phase separation in living cells~\cite{Fritsch2021}. These raise the question under what conditions local thermodynamic equilibrium approximations provide an appropriate description of biophysical processes in living cells, even though cells are  operating far from thermodynamic equilibrium. Cells host numerous active processes, the paradigmatic example is the activity of many enzymes driven by the hydrolysis of ATP such as the action of molecular motors in the cell. Such activity involves chemical events that release heat and trigger chemical changes of involved biomolecules. Standard approaches to describe such biophysical processes are based on the idea that locally temperature, pressure and other thermodynamic variables are well defined and that the non-equilibrium physics arises at larger scales by a smaller number of degrees of freedom that are maintained away from equilibrium and exhibit non-equilibrium dynamic behaviors. However this raises the question of whether and at which length and time scales such local equilibrium assumptions hold.  Our work shows that even in active systems driven by stochastic active events that release heat, there typically exist length- and time-scale regimes where fluctuations are dominated by thermal fluctuations which are consistent with local thermodynamic conditions at a local equilibrium. Therefore in these regimes local equilibrium is an accurate framework to describe the non-equilibrium dynamics that emerges at larger scales.

To describe non-equilibrium conditions at small scales, we focus our work on fluctuations of heat, which can be either active, i.e. when a chemical event takes place far from equilibrium or passive when they are related to local equilibrium. To characterize active heat fluctuations we have introduced  a stochastic Poisson field theory. Because of the Poissonian fluctuations, higher cumulants do not vanish and can be calculated explicitly. Furthermore, we find that correlation functions exhibit power-law behaviors in space and time which correspond to out-of-equilibrium critical behaviors without the need to tune parameters to a critical point. In frequency space, the noise spectrum decreases as a power-law. Notably, in two spatial dimensions, we find $1/f$-noise. Poisson white noise is usually studied for  discrete variables~\cite{FeynmanHibbs1965,Hanggi1978A,Hanggi1979,Hanggi1978,Hanggi1980}. Here we present a  continuum theory in space and time with Poisson statistics. 

The approach presented in this work also used some approximations. First, we have assumed that the chemical reaction are all  the same type, and second, have neglected  non-linearities in the heat conduction equation. While it is straightforward to extend our analysis to include several types of reactions, the problem of non-linear heat transfer is more challenging and arises already in the absence of activity~\cite{Ozisic1993}. 
All these limitations present however directions for future developments.

Our analysis and the results presented here can also be applied to other active systems.  Examples are particles propelled by chemical fuel, transport processes that generate heat via stochastic molecular events.
The description of fluctuations in such systems in general requires accounting for the coupling to mass or momentum transfer suggesting more complex behaviors of the corresponding correlation functions~\cite{Marchetti2013}.

We have shown that a field theory with a stochastic Poisson noise provides a framework that is well-suited to describe out-of-equilibrium fluctuations. Moreover, with the inclusion of sinks, such a theory provides a  new description for birth and death processes and could potentially be extended to more general  reaction-diffusion problems. Another direction to investigate is going beyond the free theory by adding interactions. Such interactions  might be relevant, for example, to describe a temperature gradient or could be taken into account when describing chemical reactions.

\section*{Acknowledgments}
The authors thank Peter H\"anggi (University of Augsburg), Uwe T\"auber (Virginia Tech) and Pierre Gaspard (Universit\'e Libre de Bruxelles) for stimulating discussions and comments on the manuscript.

\begin{appendices}
\begin{widetext}

\appendix
\section{Statistics of active fluctuations}
\label{ap:gfdtA}
In this appendix, we review the statistical properties of the temperature fluctuations $\delta T_\text{A}$ due to the Poisson white noise. For the sake of generality, a system of $d$ spatial dimensions is considered.
The temperature fluctuation $\delta T_\text{A}$ given in Eq.~\eqref{eq:deltaTA} is a stochastic field, as it depends on the number of events $n(t,t_0)$  obeying Poisson statistics and the time and location of each metabolic event, whose probability distribution is given by $\lambda(\vect{x},t)d\vect{x}dt/\Lambda$ where $\Lambda$ is the normalization, defined as
\begin{align}
	\Lambda\equiv\int_{t_0}^tdt^\prime\int_Vd^d\vect{x}\ \lambda(\vect{x},t^\prime)\;,
\end{align}
which corresponds to the average number of events $\langle n(t,t_0)\rangle$.
The moment generating functional for $\delta T_\text{A}$ is defined as
 \begin{align}
	\Phi_{t,V}[u]&\equiv\left\langle\exp\left[i\int_{t_0}^tdt^\prime\int_Vd^d\vect{x}\ u(\vect{x},t^\prime)\delta T_\text{A}(\vect{x},t^\prime)\right]\right\rangle\notag\\
	&=\exp\left[-i\int_{t_0}^tdt^\prime\int_Vd^d\vect{x}\ u(\vect{x},t^\prime)\bar{T}(\vect{x})\right]\left\langle\prod_{i=1}^{n(t,t_0)}\exp\left[i\frac{h_0}{\rho c_{P}}\int_{t_0}^tdt^\prime\int_Vd^d\vect{x}\ u(\vect{x},t^\prime)G(\vect{x},t^\prime|\vect{x}_i,t_i)\right]\right\rangle\notag\;.
 \end{align}
In order to evaluate the ensemble average we generalize the method of Ref.~\cite{FeynmanHibbs1965} to include the spatial dependence of the stochastic variable. An integration over the probability density of the individual events and an average over the Poisson distribution $P(n(t,t_0)=n)=\frac{1}{n!}e^{-\Lambda}\Lambda^n$ are performed. We obtain
 \begin{align}
    &\left\langle\prod_{i=1}^{n(t,t_0)}\exp\left[i\frac{h_0}{\rho c_{P}}\int_{t_0}^tdt^\prime\int_Vd^d\vect{x}\ u(\vect{x},t^\prime)G(\vect{x},t^\prime|\vect{x}_i,t_i)\right]\right\rangle\notag\\
    &=\sum_{n=0}^{\infty}\frac{1}{n!}e^{-\Lambda}\Lambda^n\prod_{i=1}^{n}\frac{1}{\Lambda}\int_{t_0}^tdt_i\int_Vd^d\vect{x}_i\ \lambda(\vect{x}_i,t_i)\exp\left[i\frac{h_0}{\rho c_{P}}\int_{t_0}^tdt^\prime\int_Vd^d\vect{x}\ u(\vect{x},t^\prime)G(\vect{x},t^\prime|\vect{x}_i,t_i)\right]\notag\\
	&=\sum_{n=0}^{\infty}\frac{1}{n!}e^{-\Lambda}\left(\int_{t_0}^tdt^{\prime\prime}\int_Vd^d\vect{x}'\lambda(\vect{x}',t^{\prime\prime})\exp\left[i\frac{h_0}{\rho c_{P}} \int_{t_0}^tdt^\prime\int_Vd^d\vect{x}\ u(\vect{x},t^\prime)G(\vect{x},t^\prime|\vect{x}^\prime,t^{\prime\prime})\right]\right)^{n}\notag\\
	&=\exp\left\{-\int_{t_0}^tdt^{\prime\prime}\int_Vd^d\vect{x}'\lambda(\vect{x}',t^{\prime\prime})\left(1-\exp\left[i\frac{h_0}{\rho c_{P}} \int_{t_0}^tdt^\prime\int_Vd^d\vect{x}\ u(\vect{x},t^\prime)G(\vect{x},t^\prime|\vect{x}^\prime,t^{\prime\prime})\right]\right)\right\}\;.
 \end{align}
The moment generating functional is therefore
 \begin{align}
 	\Phi_{t,V}[u]&=\exp\left\{-\int_{t_0}^tdt^{\prime\prime}\int_Vd^d\vect{x}'\ iu(\vect{x}',t^{\prime\prime})\bar{T}(\vect{x}')+\lambda(\vect{x}',t^{\prime\prime})\left(1-\exp\left[i\frac{h_0}{\rho c_{P}} \int_{t_0}^tdt^\prime\int_Vd^d\vect{x}\ u(\vect{x},t^\prime)G(\vect{x},t^\prime|\vect{x}^\prime,t^{\prime\prime})\right]\right)\right\}\;.
 \end{align}
The cumulant generating functional $\Psi_{t,V}[u]\equiv \ln \Phi_{t,V}[u]$ reads
 \begin{align}
	\Psi_{t,V}[u]&=\int_{t_0}^tdt^{\prime\prime}\int_Vd^d\vect{x}'\ \lambda(\vect{x}',t^{\prime\prime})\left(\exp\left[i \frac{h_0}{\rho c_{P}}\int_{t_0}^tdt^\prime\int_Vd^d\vect{x}\ u(\vect{x},t^\prime)G(\vect{x},t^\prime|\vect{x}^\prime,t^{\prime\prime})\right]-1\right)- iu(\vect{x}',t^{\prime\prime})\bar{T}(\vect{x}')\;.
 \end{align}
The mean value vanishes and the $m$-point cumulant, for $m>1$, is
\begin{align}
\left\langle{\delta T_\text{A}(\vect{x}_1,t_1)}...{\delta T_\text{A}(\vect{x}_m,t_m)}\right\rangle_\text{c} &= (-i)^m\left.\frac{\delta^m}{\delta u(\vect{x}_1,t_1)...\delta u(\vect{x}_m,t_m)}\Psi_{t,V}[u]\right|_{u=0}\notag\\
&=\left(\frac{h_0}{\rho c_{P}}\right)^m\int_{t_0}^tdt^\prime\int_Vd^d\vect{x}\ \lambda(\vect{x},t^\prime)G(\vect{x}_1,t_1|\vect{x},t^\prime)...G(\vect{x}_m,t_m|\vect{x},t^\prime)\;,
\end{align}
where in $d$ spatial dimension, the heat kernel $G$ reads
\begin{align}
	G(\vect{x},t|\vect{x}',t')&=\frac{\theta(t-t')}{\left[4\pi\alpha\left|t-t'\right|\right]^{\frac{d}{2}}}\exp\left[-\frac{(\vect{x}-\vect{x}')^2}{4\alpha|t-t'|}\right]\;.
\end{align}
Let us also highlight the derivation of the $2$-point cumulant for  a constant rate per unit volume $\lambda(\vect{x},t)=\lambda$ and sending the limit of spatial integration to infinity. We have
\begin{align}
\left\langle{\delta T_\text{A}(\vect{x}_1,t_1)}{\delta T_\text{A}(\vect{x}_2,t_2)}\right\rangle_\text{c} &=\lambda\left(\frac{h_0}{\rho c_{P}}\right)^2\int_{t_0}^tdt^\prime\int_{\mathbb{ R}^{d}}d^d\vect{x}\ G(\vect{x}_1,t_1|\vect{x},t^\prime)G(\vect{x}_2,t_2|\vect{x},t^\prime)\notag\\
&=\frac{\lambda h_0^2}{\left(4\pi\alpha\right)^{d}\rho^2 c_{P}^2}\int_{t_0}^t dt^\prime\frac{\theta(t_1-t^\prime)\theta(t_2-t^\prime)}{\left[\left(t_1-t^\prime\right)\left(t_2-t^\prime\right)\right]^{\frac{d}{2}}}\exp\left[-\frac{(\vect{x}_1-\vect{x}_2)^2}{4\alpha\left(t_1+t_2-2t^\prime\right)}\right]\notag\\
&\cdot\int_{\mathbb{ R}^d}d^d\vect{x}\ \exp\left\{-\frac{\left(t_1+t_2-2t^\prime\right)}{4\alpha(t_1-t^\prime)(t_2-t^\prime)}\left[\vect{x}-\frac{(t_2-t^\prime)\vect{x}_1+(t_1-t^\prime)\vect{x}_2}{t_1+t_2-2t^\prime}\right]^2\right\}\;,
\end{align}
assuming that $\vect{x}_1\neq\vect{x}_2$ and $t_1\neq t_2$ and using that
\begin{align}
	\frac{(\vect{x}_1-\vect{x})^2}{t_1-t^\prime}+\frac{(\vect{x}_2-\vect{x})^2}{t_2-t^\prime} & = \frac{\left(t_1+t_2-2t^\prime\right)}{(t_1-t^\prime)(t_2-t^\prime)}\left(\vect{x}-\frac{(t_2-t^\prime)\vect{x}_1+(t_1-t^\prime)\vect{x}_2}{t_1+t_2-2t^\prime}\right)^2+\frac{(\vect{x}_1-\vect{x}_2)^2}{\left(t_1+t_2-2t^\prime\right)}\;.
\end{align}
The integral over $d^d\vect{d}$ is a $d$-dimensional Gaussian integral, we obtain
\begin{align}
\left\langle{\delta T_\text{A}(\vect{x}_1,t_1)}{\delta T_\text{A}(\vect{x}_2,t_2)}\right\rangle_\text{c} &=\frac{\lambda h_0^2}{\left(4\pi\alpha\right)^{\frac{d}{2}}\rho^2c_{P}^2}\int_{t_0}^{\text{min}(t_1,t_2)}dt^\prime\ \frac{\exp\left[-\frac{(\vect{x}_1-\vect{x}_2)^2}{4\alpha\left(t_1+t_2-2t^\prime\right)}\right]}{\left(t_1+t_2-2t^\prime\right)^{\frac{d}{2}}}\;,
\end{align}
the change of variable $s\equiv{(\vect{x}_1-\vect{x}_2)^2}/{4\alpha (t_1+t_2-2t^\prime)}$ gives
\begin{align}
	\left\langle{\delta T_\text{A}(\vect{x}_1,t_1)}{\delta T_\text{A}(\vect{x}_2,t_2)}\right\rangle_\text{c} =&\frac{\lambda h_0^2}{2\left(4\pi\alpha\right)^{\frac{d}{2}}\rho^2c_{P}^2}\left[\frac{(\vect{x}_1-\vect{x}_2)^2}{4\alpha}\right]^{1-\frac{d}{2}}\int_{\frac{(\vect{x}_1-\vect{x}_2)^2}{4\alpha(t_1+t_2-2t_0)}}^{\frac{(\vect{x}_1-\vect{x}_2)^2}{4\alpha|t_1-t_2|}}ds\ {s}^{\frac{d}{2}-2}e^{-s}\notag\\
	=&\frac{\lambda h_0^2}{2\left(4\pi\alpha\right)^{\frac{d}{2}}\rho^2c_{P}^2}\left[\frac{(\vect{x}_1-\vect{x}_2)^2}{4\alpha}\right]^{1-\frac{d}{2}}\left[\gamma\left(\frac{d}{2}-1,\frac{(\vect{x}_1-\vect{x}_2)^2}{4\alpha|t_1-t_2|}\right)\right.\notag\\
	&\left.-\gamma\left(\frac{d}{2}-1,\frac{(\vect{x}_1-\vect{x}_2)^2}{4\alpha(t_1-t_0+t_2-t_0)}\right)\right]\;,
\end{align}
where $\gamma(s,x)$ is the lower incomplete gamma function. In three spatial dimensions, we obtain 
\begin{align}
	\left\langle{\delta T_\text{A}(\vect{x}_1,t_1)}{\delta T_\text{A}(\vect{x}_2,t_2)}\right\rangle_{c} &=\frac{\lambda h_0^2}{8\pi\alpha \rho^2c_{P}^2|\vect{x}_1-\vect{x}_2|}\left[\text{Erf}\left(\frac{|\vect{x}_1-\vect{x}_2|}{2\sqrt{\alpha|t_1-t_2|}}\right)-\text{Erf}\left(\frac{|\vect{x}_1-\vect{x}_2|}{2\sqrt{\alpha(t_1+t_2-2t_0)}}\right)\right]\;,
\end{align}
using that $\gamma(1/2,x)=\sqrt{\pi}\text{Erf}(\sqrt{x})$.

\section{Power spectra of active fluctuations}
\label{eq:1overfnoise}

In $d$-spatial dimensions, the two-point cumulant $\left\langle{\delta \hat{T}_\text{A}(\vect{x}_1,\omega_1)}{\delta \hat{T}_\text{A}(\vect{x}_2,\omega_2)}\right\rangle_\text{c}$ is obtained by taking the Fourier transform of Eq.~\eqref{eq:npointcumulantfourierspace}  into real space, which gives
\begin{align}
	\left\langle{\delta \hat{T}_\text{A}(\vect{x}_1,\omega_1)}{\delta \hat{T}_\text{A}(\vect{x}_2,\omega_2)}\right\rangle_\text{c}&=\lambda\left(\frac{h_0}{\alpha \rho c_{P}}\right)^2(2\pi)\delta(\omega_1+\omega_2)\int \frac{d^dq}{(2\pi)^d}\frac{e^{i\vect{q}.|\vect{x}_1-\vect{x}_2|}}{q^4+\left(\frac{\omega_1}{\alpha}\right)^2}\;.
\end{align}
The spectral density is then defined as
\begin{align}
	S_{\text{A}}(|\vect{x}|,\omega)\equiv \lambda\left(\frac{h_0}{\alpha \rho c_{P}}\right)^2\int \frac{d^dq}{(2\pi)^d}\frac{e^{i\vect{q}.|\vect{x}|}}{q^4+\left(\frac{\omega}{\alpha}\right)^2}\;.
\end{align}
For spatial dimension $d=3,2,1$ we obtain for spectral densities:
\begin{align}
	\left.S_{\text{A}}(|\vect{x}|,\omega)\right|_{d=3}&=\lambda\left(\frac{ h_0 }{\alpha\rho c_{P}}\right)^2\left(\frac{\omega}{\alpha}\right)^{-\frac{1}{2}}\frac{1}{4\sqrt{2}\pi}\frac{\sin\sqrt{\frac{\omega|\vect{x}|^2}{2\alpha}}}{\sqrt{\frac{\omega|\vect{x}|^2}{2\alpha}}}e^{-\sqrt{\frac{\omega|\vect{x}|^2}{2\alpha}}}\;,\label{eq:spectal_densities3}\\
	\left.S_{\text{A}}(|\vect{x}|,\omega)\right|_{d=2}&=-\lambda\left(\frac{ h_0 }{\alpha\rho c_{P}}\right)^2\left(\frac{\omega}{\alpha}\right)^{-1}\frac{1}{2\pi}\textrm{kei}\sqrt{\frac{\omega|\vect{x}|^2}{\alpha}}\;,\label{eq:spectal_densities2}\\
	\left.S_{\text{A}}(|\vect{x}|,\omega)\right|_{d=1}&=\lambda\left(\frac{ h_0 }{\alpha\rho c_{P}}\right)^2\left(\frac{\omega}{\alpha}\right)^{-\frac{3}{2}}\frac{1}{2\sqrt{2}}e^{-\sqrt{\frac{\omega|\vect{x}|^2}{2\alpha}}}\left[\sin\sqrt{\frac{\omega|\vect{x}|^2}{2\alpha}}+\cos\sqrt{\frac{\omega|\vect{x}|^2}{2\alpha}}\right]\label{eq:spectal_densities1}
\end{align}
where $\textrm{kei}$ is the Kelvin function. In the limit of small frequencies $\omega \ll 2\alpha/|\vect{x}|^2$, the spectral density has the scaling behavior
\begin{align}
	S_{\text{A}}(\vect{x},\omega) = C_d\lambda\left(\frac{ h_0 }{\alpha\rho c_{P}}\right)^2 \left(\frac{\omega}{\alpha}\right)^{\frac{d}{2}-2}\;,\label{eq:spectral_density_scaling}
\end{align}
where the constant $C_d$ equals $(4\sqrt{2}\pi)^{-1}$, $-\textrm{kei}(0)(2\pi)^{-1}$ or $(2\sqrt{2})^{-1}$ for $d=3,2,1$ respectively.

\end{widetext}

\end{appendices}

\end{document}